\documentclass{article}
\usepackage{spconf,amsmath,graphicx}
\usepackage{enumitem}
\usepackage{amsmath, amssymb}
\usepackage{booktabs}
\usepackage{color}
\usepackage{tipa}
\usepackage{cite}

\title{Robust Audiovisual Speech Recognition Models with Mixture-of-Experts}
\name{Yihan Wu$^{2,1}$, Yifan Peng$^{1}$, Yichen Lu$^{1}$, Xuankai Chang$^{1}$, Ruihua Song$^{2}$, Shinji Watanabe$^{1}$}
\address{$^{1}$Carnegie Mellon University, $^{2}$Renmin University of China \\
\{yihanwu, rsong\}@ruc.edu.cn, \{yifanpen, yichenl5, xuankaic, swatanab\}@andrew.cmu.edu}
\begin{document}
%\ninept
%
\maketitle
\begin{abstract}

Visual signals can enhance audiovisual speech recognition accuracy by providing additional contextual information. Given the complexity of visual signals, an audiovisual speech recognition model requires robust generalization capabilities across diverse video scenarios, presenting a significant challenge.
In this paper, we introduce \textbf{EVA}, leveraging the mixture-of-\textbf{E}xperts for audio\textbf{V}isual \textbf{A}SR to perform robust speech recognition for ``in-the-wild'' videos. Specifically, we first encode visual information into visual tokens sequence and map them into speech space by a lightweight projection. Then, we build EVA upon a robust pretrained speech recognition model, ensuring its generalization ability. Moreover, to incorporate visual information effectively, we inject visual information into the ASR model through a mixture-of-experts module. Experiments show our model achieves state-of-the-art results on three benchmarks, which demonstrates the generalization ability of EVA across diverse video domains.

\end{abstract}
\begin{keywords}
Audiovisual speech recognition, multimodal, mixture-of-expert
\end{keywords}

\begin{figure*}[tb]
\centering
\includegraphics[width=0.9\textwidth]{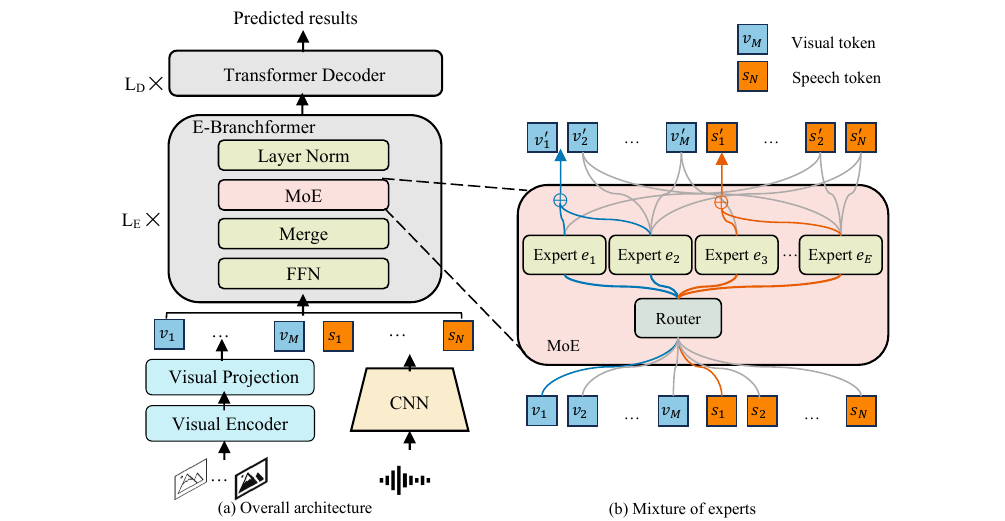}
\caption{\textbf{(a) Overall architecture of EVA. (b) The detail of the MoE layer.} For each MoE layer, only $K$ experts are activated for each token, while the other experts remain silent.}
\vspace{-1 em}
\label{fig:arch}
\end{figure*}

\section{Introduction}
\label{sec:intro}

Automatic speech recognition (ASR) systems have achieved high accuracy in standard benchmarks~\cite{ASRsurvey2021Li,E2Esurvey2024Prabhavalkar,whisper2023Radford,OWSM2023peng,OWSM2024Peng}. However when ASR models are applied to ``in-the-wild'' videos (e.g., online videos, video conferencing, TV shows, etc.), their accuracy is often degraded due to the diversity of scenes, the presence of spontaneous speech, and various noise levels~\cite{How22018Sanabria,ASRChime42018CHen,AVFormer2023Seo}. These challenges highlight the need for more robust ASR models that can effectively handle the complexities of video-based environments. Previous works~\cite{AVATAR2022Gabeur,AVFormer2023Seo} have shown that visual information can provide strong contextual information to improve the performance of ASR models.
Unlike works that focus solely on lip motion~\cite{DeepAV2022Afouras,E2eAV2021Ma,Lip2019Chung}, for ``in-the-wild'' videos, the entire visual frame contributes to ASR performance. Therefore, the model must have sufficient generalization ability to handle diverse video signals and audio backgrounds to adapt to different domains of video and corresponding audio.

% Previous audiovisual ASR models explore using unconstrained visual information from the entire visual frame to improve speech recognition models. AVATAR~\cite{AVATAR2022Gabeur} fuses visual and speech features with a multi-modal video encoder. But it cannot use the ability of the pretrained speech recognition model, which shows suboptimal performance on the out-of-domain dataset. 
Several audiovisual ASR models have been proposed to improve speech recognition performance by exploring unconstrained visual features from the entire visual frame. While AVATAR~\cite{AVATAR2022Gabeur} fuses visual and speech features with a multimodal video encoder, inferior performance is achieved on the out-of-domain dataset because the model cannot use the ability of a pretrained ASR model.
AVFormer~\cite{AVFormer2023Seo} uses an adapter to inject visual features into a pretrained ASR model. Although it achieves zero-shot generalization capability, it requires a large-scale dataset of $\left\langle \text{video, speech, text} \right\rangle$  pairs for pretraining, which demands extensive data collection. 

% We aim to improve the model's generalization ability while minimizing the requirement of large-scale multimodal pertaining data. 
This paper proposes \textbf{EVA}\footnote{The pronunciation of EVA is \textipa{[\textprimstress i\textlengthmark v@]}.}, a mixture-of-\textbf{E}xperts framework for audio\textbf{V}isual \textbf{A}SR. 
We aim to integrate unconstrained video features into existing speech recognition models to construct robust audiovisual speech recognition systems for videos across different domains.
First, to ensure generalization ability, we build EVA upon the pretrained ASR model, OWSM v3.1~\cite{OWSM2024Peng}, which is trained on large-scale public speech datasets. 
% We use OWSM instead of Whisper~\cite{whisper2023Radford} because Whisper's training data are based on web crawling, which may include our test set, potentially causing data contamination issues. 
We opt for OWSM over Whisper~\cite{whisper2023Radford} due to potential data contamination concerns. Whisper's training data, sourced through web crawling, might include our test set. 
Then, we extract frames from the input video and encode them into visual tokens using a pretrained visual encoder, resulting in strong generalization ability across different visual domains. Moreover, we design a multimodal mixture-of-experts module to integrate visual feature into the pretrained ASR model while maintaining speech recognition ability.

We summarize the main contributions as follows:
\begin{itemize}[left=0pt]
    \item We propose \textbf{EVA}, a robust audiovisual speech recognition model with mixture of experts, which shows strong generalization ability for in-the-wild videos. EVA is an open-source work that is fully reproducible\footnote{We will publicly release code in https://github.com/espnet/espnet.}.
    \item We design a novel mixture-of-experts module in EVA, enabling the model to preserve the pretrained model's speech understanding capabilities while exploring visual understanding abilities by using experts and learnable routers.
    \item We conduct extensive experiments on three datasets across different domains. These experiments demonstrate that EVA shows robust speech recognition ability for audiovisual ASR in both instructional videos and egocentric videos. It is worth noting that EVA outperforms the previous SOTA model,  AVFormer~\cite{AVFormer2023Seo}, with $\sim$400x smaller audiovisual training data.
\end{itemize}

\section{Related Works}
\label{sec:format}

% \subsection{Audiovisual speech recognition}
\textbf{Audiovisual speech recognition.}
% Recent state-of-the-art ASR models show a good performance~\cite{ASRsurvey2021Li,E2Esurvey2024Prabhavalkar}. 
Recent whisper-style ASR models~\cite{whisper2023Radford,OWSM2023peng,OWSM2024Peng} explore large-scale supervised learning to achieve good performance across different benchmarks. Based on pretrained ASR models, some works explore ASR under a multimodal scenario~\cite{AVATAR2022Gabeur,Promptwhisper2023Peng,Lip2019Chung, LLD2021Ghorbani}. With both speech and video as input, most audiovisual ASR works focus on lip motion specifically~\cite{Lip2019Chung,DeepAV2022Afouras,E2eAV2021Ma}. These methods process the pixels of the speaker's lips or use the pre-extracted lip visual features. 
% It limits the use of audiovisual ASR to scenarios where the speaker's lips are visible in the video.
This restricts audiovisual ASR to scenarios with visible speaker lips.
AVATAR~\cite{AVATAR2022Gabeur} uses the entire frame of video based on the encoder-decoder speech recognition model. The encoder performs audio-visual fusion early and is trained directly from pixels and spectrograms. 
But AVATAR only performs well on the in-domain test set, with a poor generalization ability to out-of-domain test sets.
AVFormer~\cite{AVFormer2023Seo} further explores injecting visual information into a frozen ASR model. To enhance the model's generalization ability, AVFormer is finetuned on the large-scale multimodal dataset HowTo100M~\cite{HowTo1002019Miech}. Besides, some recent works~\cite{Promptwhisper2023Peng,VisualER2023Kumar} utilize vision-language pretrained models to convert visual information into text, using the text as a prompt to enhance the accuracy of ASR models. However, this process can result in the loss of certain visual information, thereby reducing the contextual information available to the audiovisual ASR.
In this work, we improve the generalization capability systematically. We build EVA upon the robust ASR model to ensure the robust speech recognition ability. Furthermore, we design a novel mixture-of-experts module to carefully finetune the pretrained model on the audiovisual dataset to incorporate unconstrained visual information. 

% \subsection{Mixture-of-experts in multimodal learning}
\noindent
\textbf{Mixture-of-experts in multimodal learning.}
Mixture-of-Experts (MoE)~\cite{moe1991Jacobs,Deepmoe2014Eigen} is a hybrid model comprising multiple integrated sub-models, known as experts. Many previous works explore enhancing model performance with MoE in natural language processing and computer vision~\cite{sparseMoE2017Shazeer,MoELLaVA2024Lin,Deepmoe2014Eigen}. MoE module consists of several feed-forward sub-networks, with a trainable router to determine a sparse combination of these experts to use for each token.
MoE enables the dynamic allocation of data among different experts, allowing each expert to focus on its expertise and achieve model sparsity. Previous works show good performance of multimodal MoE in vision language pretrained models~\cite{EVE2024Chen,MoELLaVA2024Lin}. To the best of our knowledge, EVA first applies multimodal MoE in an audiovisual speech recognition task which achieves significant performance improvement.

\section{Proposed Method}
\label{sec:method}

The architecture of EVA is shown in Figure~\ref{fig:arch}. We improve the generalization ability of the EVA with three strategies. First, we use OWSM v3.1, a speech recognition model which is pretrained on 180k hours of public labeled speech data to ensure robust speech recognition ability.
Second, we encode unconstrained full frame visual information with the pretrained visual encoder CLIP~\cite{clip2021Radford} which has strong zero-shot capabilities. Third, we integrate visual information into OWSM v3.1 via a multimodal MoE module (as shown in Figure~\ref{fig:arch}(b)), which retains the original model's speech recognition ability while adding visual understanding ability.

\subsection{Robust speech recognition model}
\label{sec:owsm}
We start with a speech recognition backbone that achieves robust performance on traditional ASR benchmarks~\cite{LibriTTS2015Panayotov}. We use OWSM v3.1~\cite{OWSM2024Peng} that adopts an encoder-decoder architecture with a joint CTC loss for ASR targets~\cite{jointCTC2017Kim}. More specifically, OWSM v3.1  adopts an E-Branchformer~\cite{Ebranchformer2022Kim} encoder which utilizes parallel branches to capture local and global contextual information~\cite{branchformer}. 
As shown in Figure~\ref{fig:arch}(a), each E-Branchformer layer is a stack of macaron-style feed-forward networks (FFN), an enhanced merge module, and a layernorm module. 
The input waveforms are transformed into log Mel filterbanks, using a window length of 25ms and a hop length of 10ms. 
Being trained on the large-scale dataset, OWSM v3.1 shows robust performance on many traditional speech benchmarks. It supports a good generalization ability for audiovisual speech recognition tasks.

\subsection{Visual encoder}
\label{sec:visual_enc}
To encode visual information for full frames of videos across different domains, we use a pretrained visual encoder, CLIP~\cite{clip2021Radford}, to ensure generalization capability. CLIP is trained on image and text paired data based on contrastive loss, and known to have strong generalization and zero-shot capabilities. This makes our features more suited to unconstrained videos ‘in-the-wild’.
Given a sequence of frames extracted from a video, we first use the vision encoder to process input frames to obtain the visual tokens $z = [z_{1}, z_{2}, ..., z_{M}]^{T} \in \mathbb{R}^{M \times C}$, where $M$ is the number of visual tokens (equal to the number of frames) and $C$ is the token embedding dimensionality. Then, a visual projection layer is used to map $z \in \mathbb{R}^{M \times C}$ to $v \in \mathbb{R}^{M \times D}$, where $D$ matches the hidden size of the pretrained ASR model.

\subsection{Multimodal mixture-of-experts}
\label{sec:moe}
In this work, we propose the multimodal MoE module to retain the model's speech recognition capabilities while incorporating visual understanding.

Given visual tokens $v = [v_{1}, v_{2}, ..., v_{M}]^{T}\in \mathbb{R}^{M \times D}$ and speech tokens $s = [s_{1}, s_{2}, ..., s_{N}]^{T}\in \mathbb{R}^{N \times D}$, we concatenate them along the sequence dimension to form $x = [v_{1}, v_{2}, \cdots, v_{M}, s_{1}, s_{2}, ..., s_{N}]^{T}$, resulting in a combined sequence of length $N + M$. 
Then, we feed them into the encoder of the ASR model. As we introduced in Section~\ref{sec:owsm}, the encoder of OWSM is the stack of E-branchformer blocks. 
We keep all other layers except the second FFN of each block (see Figure~\ref{fig:arch}). 
Typically, an MoE layer consists of multiple FFNs and a router $f$ whose output is a sparse $E$-dimensional vector, where $E$ is the number of experts.
As shown in Figure~\ref{fig:arch}(b), we replicate the FFNs from the pretrained ASR model to form an ensemble of experts $\mathcal{E} = [e_{1}, e_{2}, \cdots, e_{E}]$, and use a linear layer as the router $f$ to predict the probability $P$ of each token being assigned to each expert:
\begin{equation}
\label{eq:moe_prob}
    P(x)_i = \frac{e^{f(x)_i}}{\sum_{j=1}^{E} e^{f_{j}(x)}},
\end{equation}
where $f(x) = W \cdot x$. Here, $W \in \mathbb{R}^{D \times E}$ represents the lightweight training parameters and $E$ is the number of experts for an MoE layer.
Therefore, each token of $x$ is processed by the top-$K$ experts with the highest probabilities. The output of the MoE layer can be written as 
\begin{equation}
\label{eq:moe_output}
\text{MoE}(x) = \sum_{i=1}^{k} P(x)_i \cdot e_{i}(x),
\end{equation}
where $e_{i}(x)$ is the output of the $i$-th expert for an input $x$.

To maintain the pretrain ASR model's speech understanding performance while incorporating visual understanding performance, proper initialization of MoE layers is important.
We first replicate the FFN weights from the pretrained ASR model as the initialization weights for the experts and then finetune the whole model on the audiovisual dataset. Finally, the model gradually transitions from a speech recognition model initialization to audiovisual ASR model with the mixture-of-experts to handle input from different modalities.

\subsection{Training objectives}
The total loss, $\mathcal{L}_{\text{total}}$, consists of an attention loss, a CTC loss, and an auxiliary loss function to encourage experts to receive roughly equal numbers of training examples.

\noindent
\textbf{Attention loss.}
The loss function of the attention model is computed from Eq.~\ref{eq:moe_prob} and Eq.~\ref{eq:moe_output} as:
\begin{align}
    \mathcal{L}_{\text{att}} & = - \ln{P_{\text{att}}(y^{\star}|x)} = - \sum_{u} \ln{P_{\text{att}}{(y_{u}^{\star}|x, y_{1:u-1}^{\star})}},
\end{align}
where $y_{1:u-1}^{\star}$ is the ground truth of the previous tokens.

\noindent
\textbf{CTC loss.}
The CTC loss to be minimized is defined as the negative log likelihood of the ground truth character sequence $Y^{\star}$, i.e.
\begin{equation}
    \mathcal{L}_{\text{CTC}} = - \ln{P_{\text{CTC}}(y^{\star}|x}).
\end{equation}

\noindent
\textbf{Auxiliary loss.} To avoid the gating network converging to a state where it always produces large weights for the same few experts, we introduce load balancing constraints on the MoE layer as the auxiliary loss~\cite{sparseMoE2017Shazeer,MoELLaVA2024Lin}. 
Assume we have a batch $\mathcal{B}$ with $T$ tokens from both video and speech.
We impose auxiliary loss into each MoE layer to encourage all experts to have equal importance as follows:
\begin{equation}
    \mathcal{L}_{\text{aux}} = E \cdot \sum_{i=1}^{E} \mathcal{F}_{i} \cdot \mathcal{G}_{i},
\end{equation}
where $\mathcal{F}_{i}$ represents the fraction of tokens processed by each expert $e_{i}$, and $\mathcal{G}_{i}$ represents the average routing probability of $e_{i}$, which can be expressed by
\begin{align}
    \mathcal{F}_{i} &= \frac{1}{T} \sum_{x \in \mathcal{B}} \mathbb{I} \{\text{argmax} P(x), i\}, \\
\mathcal{G}_{i} &= \frac{1}{T} \sum_{x \in \mathcal{B}} P(\mathbf{x})_i.
\end{align}
The total loss $\mathcal{L}_{\text{total}}$ is calculated by
\begin{equation}
    \mathcal{L}_{\text{total}} = \mathcal{L}_{\text{att}} + \alpha \cdot \mathcal{L}_{\text{CTC}} + \beta \cdot \mathcal{L}_{\text{aux}},
\end{equation}
where $\alpha$ and $\beta$ are the loss balancing coefficients.

\begin{table}[t]
\caption{Comparison to SOTA methods across different datasets. Results are reported as WER \% (lower is better). For VisSpeech and Ego4D datasets, we evaluate EVA on these two dataset without any fine-tuning. $^\dagger$ denotes these two models are trained on HowTo100M, a much larger additional audiovisual dataset with 131k hours videos.}
    \label{tab:overall}
    \centering
    \begin{tabular}{lccc}
    \toprule
        Model & How2 & VisSpeech & Ego4D \\
        \midrule  
        How2 base~\cite{How22018Sanabria} & 18.0 & -- & -- \\
        VAT~\cite{VAT2019Caglayan} & 18.0 & -- & -- \\
        MultiRes~\cite{MultiRes2017Paraskevopoulos} & 20.5 & -- & -- \\
        LLD~\cite{LLD2021Ghorbani} & 16.7 & -- & -- \\
        AVATAR~\cite{AVATAR2022Gabeur} & 15.6 & 43.4 & -- \\
        % SynesLM & 16.1 & 40.1 & -- \\
        \midrule
        AVATAR$^\dagger$~\cite{AVATAR2022Gabeur} & \textbf{9.1} & 35.7 & 92.0 \\
        AVFormer$^\dagger$~\cite{AVFormer2023Seo} & 10.2 & 16.6 & 64.8 \\
        \midrule
        % EVA-base & & & \\
        EVA-small & 9.7 & 15.8 & 59.4 \\
        EVA-medium & 9.5 & \textbf{14.8} & \textbf{57.2} \\
    \bottomrule
    \end{tabular}

\end{table}
\section{Experimental Settings}
% \subsection{Implementation details}
\textbf{Implementation details.}
As mentioned in Section~\ref{sec:owsm}, we use OWSM v3.1 as the speech recognition backbone. Based on the different sizes of OWSM v3.1, we develop EVA-small and EVA-medium based on OWSM v3.1-small and OWSM v3.1-medium, respectively. Specifically, EVA-small has 9 E-Branchformer blocks in the encoder and 9 Transformer blocks in the decoder. The hidden size is 768. EVA-medium has 18 E-Branchformer encoder blocks and 18 Transformer decoder blocks with a hidden size of 1024.
Following AVFormer, we utilize CLIP-Large~\cite{clip2021Radford} as the visual encoder and set $M = 4$ in Section~\ref{sec:visual_enc}. Therefore, 4 frames can effectively summarize the video's information. Additionally, we use a linear layer as the visual projection layer to map visual features into the speech token embedding space. When finetuning OWSM on the audiovisual dataset, we set 8 experts while active top-4 experts for each token.
We replicate the weights of the FFN to initialize each expert, while the visual projection layer is randomly initialized. We finetune EVA-small and EVA-medium from pretrained models for 10 epochs with a batch size of 64. We use 4 V100 GPUs (32GB) for EVA-small and 4 A100 GPUs (40GB) for EVA-medium respectively. 
The value of loss balancing coefficient $\alpha$ is 0.3 and $\beta$ is 0.01. 

\begin{figure*}[tb]
\centering
\includegraphics[width=0.9\textwidth]{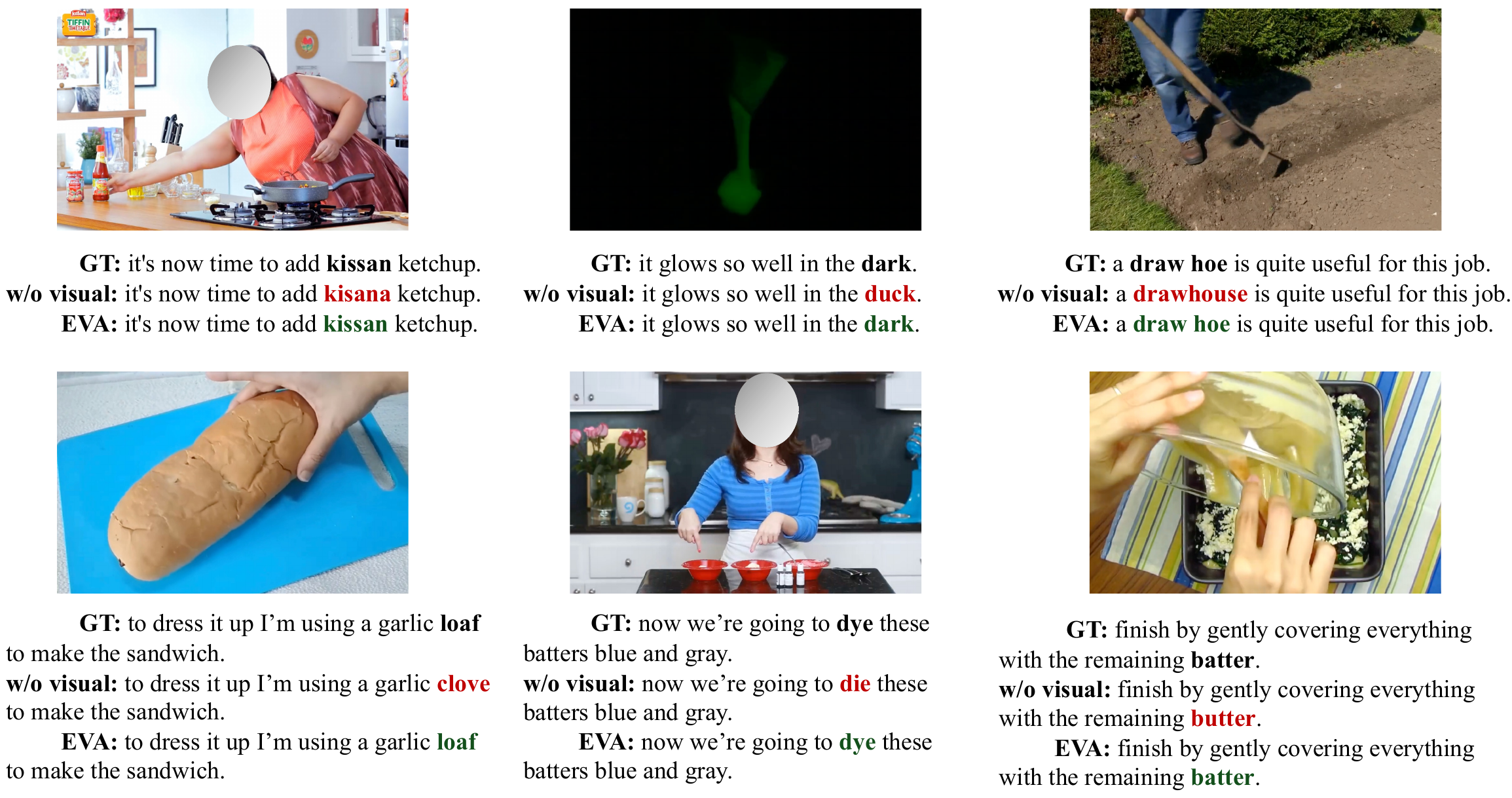}
\vspace{-1 em}
\caption{Qualitative Results. We show the ground truth text (GT), predictions from the speech-only model (w/o visual) and EVA.}
\vspace{-1 em}
\label{fig:cases}
\end{figure*}

\begin{table*}[t]
\caption{Ablation study results across different datasets, reported as WER\% (lower is better). Blue digits indicate the relative decline in model performance (relative increase in WER). "A" and "V" refer to audio and visual, respectively.}
    \label{tab:ablation}
    \centering
    \begin{tabular}{l|c|c|ccc}
    \toprule
        Model & Modality & Finetuing dataset &How2 & VisSpeech & Ego4D \\
        \midrule  
        EVA-medium & A+V & How2 & 9.5 & 14.8 & 57.2 \\
        \quad - Mixture-of-experts & A+V & How2 & 9.8 {\color{blue}(-3.2\%)} & 15.3 {\color{blue}(-3.4\%)} & 59.9 {\color{blue}(-4.7\%)}\\
        \qquad - Visual & A & How2 & 10.3 {\color{blue}(-8.4\%)} & 16.6 {\color{blue}(-12.2\%)} & 70.6 {\color{blue}(-23.4\%)}\\
        \quad \qquad - Finetune & A & -- & 22.2 {\color{blue}(-133.7\%)} & 23.9 {\color{blue}(-61.5\%)} & 71.8 {\color{blue}(-25.5\%)}\\
    \bottomrule
    \end{tabular}
    \vspace{-1em}
\end{table*}
\noindent 
\textbf{Datasets.}
We finetune the model on How2 dataset and evaluate EVA on three datasets across different domains, i.e. instructional videos datasets How2 and VisSpeech, and egocentric videos dataset Ego4D.

\begin{itemize}[left=0pt]
    \item How2~\cite{How22018Sanabria} is an instructional video dataset designed for multimodal understanding. Following AVFormer~\cite{AVFormer2023Seo}, we use the 300-hour version of How2. These videos are segmented into short clips, averaging 5.8 seconds each.
    % Each clip is accompanied by a user-uploaded transcript, averaging 20 words. 
    The dataset is divided into training (184,949 clips), validation (2,022 clips), and test (2,305 clips) sets.
    \item VisSpeech~\cite{AVATAR2022Gabeur} is an audiovisual ASR testset sampled from the HowTo100M dataset. It contains 508 clips with manually annotated transcripts. VisSpeech uses a video-text similarity model to ensure high audio-visual correspondence.
    \item Ego4D~\cite{Ego4D2022Grauman} is an egocentric daily-life activity video dataset. We utilize the audiovisual diarization benchmark from the Ego4D challenge. Our model is evaluated on the validation set (51 clips) which has ground truth annotations. We segment 5-minute long videos into shorter clips for analysis. It should be noted that compared to How2 and VisSpeech with instructional videos, the Ego4D dataset consists of more noisy and spontaneous videos across different domains, increasing the difficulty of speech recognition.

\end{itemize}

% \begin{figure}[!tb]
% \centering
% \includegraphics[width=0.45\textwidth]{latex/fig3_moe.pdf}
% \caption{Distribution of modalities across different experts. Different colors represent different experts.}
% \label{fig:arch}
% \end{figure}

\label{sec:exp}

\section{Results}

\textbf{Comparison with SOTA models}
We compare EVA with SOTA models on three datasets in Table~\ref{tab:overall}. Compared to models trained on the same dataset (How2 base, VAT, MultiRes, LLD and AVATAR), EVA achieves the best performance on all three datasets. AVATAR$^{\dagger}$ shows the best performance on How2 because it is trained on HowTo100M, a large-scale audiovisual dataset with 131k hours videos. However, as AVATAR fuses visual features and speech features in a deep fusion method, it shows poor performance on other datasets, i.e. VisSpeech and Ego4D. Compared with baseline models, EVA shows consistently better results across different domain datasets, especially on the out-of-domain egocentric dataset Ego4D. It demonstrates the robustness and generalization ability of EVA.

\noindent
\textbf{Ablation studies}
In this section, we conduct ablation studies to verify the
effectiveness of each component in EVA. In
Table ~\ref{tab:ablation}, we have the following observations:
\begin{itemize}[left=0pt]
    \item Removing the MoE module leads to a recognition accuracy drop in all three datasets, especially for the out-of-domain dataset Ego4D. It verifies that employing the MoE module can improve model generalization ability.
    \item Discarding visual features impairs recognition accuracy significantly. This removal results in the largest drop in Ego4D, with a 23.4\% relative WER increase. This demonstrates the necessity of incorporating visual information to improve ASR accuracy in diverse and noisy datasets.
    \item The last row shows that removing the finetuning stage leads to the largest performance drop compared with other settings. It highlights the importance of finetuning the pretrained ASR model on the audiovisual dataset due to the domain mismatch.  

\end{itemize}

\noindent
\textbf{Qualitative analysis}
To illustrate how the EVA improves ASR with visual information, we present a few examples in Figure~\ref{fig:cases}. Experiments show that speech recognition models are prone to errors with nouns, whereas visual information can provide additional context, especially when there is a strong correlation between the visual content and these specific words. For example, visual information helps the model recognize infrequent words such as brand names (row 1 column 1).
Additionally, visual information helps determine the correct word among similarly pronounced words (row 1 column 2 and row 2 column 3) or homophones (row 2 column 2) based on visual cues.

\section{Conclusion}
\label{sec:conclusion}
In this paper, we develop EVA, a robust audiovisual speech recognition model for in-the-wild videos across different domains. First, we encode visual information into a sequence of visual tokens. Then, we build the audiovisual ASR model upon a robust pretrained speech recognition model, ensuring its generalization ability. Moreover, we inject visual information into the ASR model through a mixture-of-experts module. We show our model achieves state-of-the-art results on three benchmarks, which shows the generalization ability of EVA across diverse video domains.
For future work, we will explore parameter-efficient fine-tuning for audiovisual ASR. Also, we will explore models with better generalization ability, which can achieve higher accuracy for general videos. 

\section{ACKNOWLEDGMENTS}
\label{sec:ack}

Experimnets of this work used the Bridges2 system at PSC
and Delta system at NCSA through allocations CIS210014
and IRI120008P from the Advanced Cyberinfrastructure Coordination Ecosystem: Services \& Support (ACCESS) program.

\newpage
\bibliographystyle{IEEEbib}
\bibliography{refs}

\end{document}